\documentclass[acmsmall,screen,nonacm]{acmart}
\usepackage{multirow} 
\usepackage{array}    
\usepackage{graphicx} 
\usepackage{adjustbox}
\usepackage{makecell}
\usepackage{tcolorbox}
\usepackage{tabularx}

\newcolumntype{C}[1]{>{\centering\arraybackslash}m{#1}}
\usepackage{indentfirst}
\usepackage{algorithmic}
\usepackage{graphicx}
\usepackage{textcomp}
\usepackage{xcolor}
\usepackage{comment}
\usepackage{hyperref}
\usepackage{bm}
\usepackage{enumitem}
\usepackage{appendix}

\usepackage{amsmath}
\usepackage{lipsum}
\usepackage{multirow}
\usepackage[justification=centering]{caption}
\usepackage{caption}
\usepackage{url}
\usepackage{listings}
\usepackage{subfigure}
\usepackage{array}
\usepackage{pifont}

\begin{document}

\title{Predicting Developer Acceptance of AI-Generated Code Suggestions}

\author{Jing Jiang}
\email{jiangjing@buaa.edu.cn}

\author{Liehao Li}
\email{liehao@buaa.edu.cn}

\author{Jinyun Hou}
\email{zy2206235@buaa.edu.cn}

\author{Xin Tan}
\authornote{Corresponding author.}
\email{xintan@buaa.edu.cn}

\author{Li Zhang}
\email{lily@buaa.edu.cn}

\affiliation{%
  \department{State Key Laboratory of Complex \& Critical Software Environment}
  \department{School of Computer Science and Engineering}
  \institution{Beihang University}
  \city{Beijing}
  \country{China}
}

\renewcommand{\shortauthors}{Jiang et al.}

\begin{abstract}

AI-assisted programming tools are widely adopted, yet their practical utility is often undermined by undesired suggestions that interrupt developer workflows and cause frustration. While existing research has explored developer-AI interactions when programming qualitatively, a significant gap remains in quantitative analysis of developers’ acceptance of AI-generated code suggestions, partly because the necessary fine-grained interaction data is often proprietary. To bridge this gap, this paper conducts an empirical study using 66,329 industrial developer-AI interactions from a large technology company. We analyze features that are significantly different between accepted code suggestions and rejected ones. We find that accepted suggestions are characterized by significantly higher historical acceptance counts and ratios for both developers and projects, longer generation intervals, shorter preceding code context in the project, and older IDE versions. Based on these findings, we introduce \emph{CSAP} (Code Suggestion Acceptance Prediction) to predict whether a developer will accept the code suggestion before it is displayed. Our evaluation of \emph{CSAP} shows that it achieves the accuracy of 0.973 and 0.922 on imbalanced and balanced dataset respectively. Compared to a large language model baseline and an in-production industrial filter, \emph{CSAP} relatively improves the accuracy by 12.6\% and 69.5\% on imbalanced dataset, and improves the accuracy by 87.0\% and 140.1\% on balanced dataset. Our results demonstrate that targeted personalization is a powerful approach for filtering out code suggestions with predicted rejection and reduce developer interruption. To the best of our knowledge, it is the first quantitative study of code suggestion acceptance on large-scale industrial data, and this work also sheds light on an important research direction of AI-assisted programming.

\end{abstract}

\ccsdesc[500]{Software and its engineering~Automatic programming~Empirical software validation}

\keywords{Accepted Code Suggestion, AI-Assistant Programming, Code Generation}

\maketitle

\section{Introduction}
\label{sec:intro}

AI-assisted programming tools powered by large language models, such as GitHub Copilot and Cursor, are aimed at providing programmers with code suggestions to help improve their productivity ~\cite{bakal2025experience,zhou2025exploring,dresselhaus2025field,pandey2024transforming}. These systems usually operate by displaying the suggestion as ghost text—a grayed-out code suggestion inline inside the IDE, which programmers can accept or reject. However, not all the code suggestions generated by these state-of-the-art systems are helpful to the developers. For instance, one study on GitHub Copilot found that developers rejected approximately 70\% of the suggestions shown to them \cite{ziegler2022productivity}. This high rate of unhelpful suggestions poses a significant threat to the primary goal of these tools. When a suggestion is rejected, it can disrupt a developer's cognitive flow, forcing them to pause their work to review and discard the irrelevant code, which wastes valuable time and leads to frustration \cite{barke2023grounded,vaithilingam2022expectation,guglielmi2025copilot,simkute2025ironies,hellendoorn2019code,li2021toward}. This highlights a critical need to understand the underlying drivers of suggestion acceptance.

The difference between accepted code suggestions and rejected code suggestions is not yet well understood. While existing research has explored developer interaction with these tools qualitatively ~\cite{tan2024far,barke2023grounded,liang2024large,tu2014localness,pearce2025asleep,suryavanshi2025integrating}, a significant gap remains in the quantitative analysis of suggestion acceptance. One reason is that much of the fine-grained interaction data resides within company-internal environments and is not publicly accessible. Moreover, the log data that can typically be captured from IDEs—such as acceptance or rejection events—are insufficient on their own to explain or predict developers' decisions. Consequently, the problem of predicting the acceptance of AI-generated code suggestions remains a relatively underexplored area of research. 
Without acceptance prediction for code suggestions,
AI-assisted programming tools directly provide code suggestions to developers,
and some suggestions are rejected,
which may negatively impact the developer experience and reduce their willingness to use the code completion tool. Therefore, building a model to predict the acceptance of code suggestions is crucial for optimizing developer-AI collaboration in software engineering.
If code suggestions are predicted to be rejected,
they can be removed without disrupting developers.

In this paper, we conduct an empirical study about accepted code suggestions. 
We collect 66,329 developer-AI interaction logs from our industry partner, encompassing every accepted and rejected suggestion, along with associated code context, developer characteristics, and project environments. Based on this company-internal dataset, we posit the following research questions:

\begin{itemize}
    \item \textbf{RQ1: What are the significant features of the accepted code suggestions?}   
\end{itemize}

Refer to previous works and information provided by our industry partner,
we design candidate features from three dimensions: developer habit, project preference, and in-situ context.
Developer habit features and project preference features analyze historical characteristics of a developer or a project, respectively.
In-situ context features capture a snapshot of the immediate circumstances surrounding a single code suggestion. 
We perform a significance analysis and identify important features which accepted code suggestions
and rejected code suggestions have significantly different values.
We find that: At the developer level, accepted code suggestions are characterized by the developer's significantly higher historical acceptance count and ratio, and longer generation intervals; At the project level, they are characterized by the project's significantly higher historical acceptance count and ratio, and a shorter preceding code context;
Furthermore, accepted code suggestions have older IDE versions.

\begin{itemize}
    \item \textbf{RQ2: How can we design an effective prediction method to determine whether a developer will accept a code suggestion?}
\end{itemize}

Based on features which are significantly different between accepted and rejected code suggestions in RQ1,
we perform the correlation analysis and derive the feature set for prediction.
We use a simple neural network with class-balanced binary cross-entropy loss function to build \emph{CSAP} (Code Suggestion Acceptance Prediction), and predict whether a developer will accept a code suggestion before it is displayed.
The evaluation of \emph{CSAP} shows that it achieves the accuracy of 0.973 and 0.922 on imbalanced and balanced dataset respectively. Compared to the direct LLM call (Qwen2.5-Coder-32B) and the Circuit Breaker model (in-production industrial filter),
 \emph{CSAP} improves the accuracy by 12.6\% and 69.5\% on imbalanced dataset,
 and improves the accuracy by 87\% and 140.1\% on balanced dataset. Feature analysis reveals that the most critical predictors, stemming from in-situ context, developer habits, are the \textit{in-situ\_IDE\_version}, and \textit{developer\_accepted\_ratio}. These features highlight the importance of tailoring predictions to a developer's immediate environment and past behavior.

This paper makes the following contributions:
\begin{itemize}
    \item We conduct a first large-scale quantitative study using fine-grained enterprise IDE interaction data and identify important features of accepted code suggestions.
    \item We propose and evaluate \emph{CSAP}, a personalized code acceptance prediction method, which can be used to filter out code suggestions with rejected prediction and reduce developer interruption. 
\end{itemize}

\section{BACKGROUND AND RELATED WORK}
\label{sec:background}

\subsection{AI-Assisted Programming}

\subsubsection{Developer Needs in AI-Assisted Programming}

The recent proliferation of large language models~\cite{guo2025deepseek,yang2025qwen3,comanici2025gemini} has spurred the development of powerful AI-assisted programming tools that leverage these models for code generation~\cite{zheng2023codegeex,li2022competition,gao2025trae,anthropic2024claude3_7}. Consequently, understanding developer needs has become a critical area of research to ensure these tools are effective and well-adopted. A study by Davila et al. ~\cite{davila2024industry} shows that a primary motivation for using these tools is to write better or alternative solutions to programming problems. However, effectiveness extends beyond just the quality of the output. To maintain a seamless development workflow, it is crucial that these tools avoid disruption, for instance, by filtering out low-confidence suggestions as argued by Ciniselli et al. ~\cite{ciniselli2023source}. Furthermore, to truly accelerate the development process, developers expect generated code to align with their personal coding styles and preferences~\cite{liang2024large,ciniselli2023source,tan2024far,wang2023practitioners,kyaw2018proposal,bird2022taking,brown2024identifying}.

\subsubsection{Factors Influencing Developers' Acceptance of Code Suggestions}

As AI-assisted programming tools proliferate, a growing body of research has examined the developer-experience determinants of developers' acceptance of AI-generated code suggestions. Tan et al. \cite{tan2024far} conducted interviews with professional developers and synthesized factors shaping adoption and trust in intelligent tools. Barke et al. \cite{barke2023grounded} analyzed interaction patterns around GitHub Copilot, characterizing how developers negotiate, edit, and reject suggestions during day-to-day use. Liang et al. \cite{liang2024large} surveyed 410 developers about Copilot and related tools, cataloging common reasons for rejection. Collectively, these studies provide qualitative foundations for understanding acceptance behavior. However, large-scale quantitative evidence, especially fine-grained in-IDE telemetry that captures real decision signals, remains limited. This gap motivates our study, which utilizes enterprise developer-AI interaction logs to quantify which factors are significantly different between accepted code suggestions and rejected code suggestions.

\subsection{Predictive Modeling of Developer Decision}
\label{subsec:predictive modeling of developer decision}

Predictive modeling of developer behavior is a long-standing topic in software engineering \cite{mozannar2024reading,bao2017will,kamei2012large,nagappan2005use}. By contrast, work that directly predicts whether developers will accept AI-generated code suggestions remains scarce, in part because enterprise IDE deployments rarely expose fine-grained, real-time usage logs of developers' coding actions that record actual acceptance decisions.

A related thread studies code-completion quality estimation, such as FrugalCoder (QEBC) \cite{sun2024dontcompleteitpreventing}, which builds lightweight early-rejection estimators using offline quality metrics computed from prompt-completion pairs. These approaches focus on intrinsic completion quality and efficiency rather than on personalized acceptance, and they typically do not model developer- or project-specific context.

We also draw on adjacent research that predicts pull-request (PR) outcomes, which similarly models developer decisions about code artifacts but in asynchronous review workflows rather than real-time IDE interactions. Representative methods include CARTESIAN \cite{azeem2020action}, which aggregates features across projects, PRs, contributors, and reviewers and trains tree-based learners; BatchRL and ChatRL ~\cite{joshi2024comparative}, which formulate decision making over multi-dimensional PR attributes (e.g., files changed, project signals, discussion sentiment) by framing outcome prediction as a sequential decision-making problem that is solved using reinforcement learning; PredCR \cite{islam2022early}, which integrates reviewer-, contributor-, project-, text-, and code-level features to predict merge/abandon; and Khatoonabadi et al. \cite{khatoonabadi2023wasted}, who analyze 16 features spanning PRs, contributors, review process, and projects via mixed quantitative–qualitative methods. While these works target PR acceptance, their feature-design principles—combining artifact, developer, and process signals—inform our selection of code-, developer-, and project-level features and our quantitative evaluation of feature significance for the acceptance-prediction task.

    \section{RQ1: Significant Features of the Accepted Code Suggestions}
    \label{sec:rq1}
    
    In this section, leveraging data from our industry partner, we identify significant features of the accepted code suggestions. Specifically, in RQ1 we: (1) describe the rationale of feature design; (2) propose candidate features; and (3) test their significance to find features which accepted code suggestions and rejected code suggestions have significantly different values.

    \subsection{Rationale of Feature Design}
    \label{sec:literature review and developer survey}
    
    \subsubsection{Circuit Breaker (abbreviated as Cbk)}
    This study is conducted in partnership with a leading global technology company specializing in software and systems development.
    We investigate their internally deployed, LLM-driven IDE, which provides real-time code completion and generation. 
    This AI-assisted programming tool integrates with mainstream IDEs like VS Code and has seen significant adoption, with over 10,000 installations across numerous engineering teams. This large-scale deployment provides the real-world setting for our study. AI-generated code suggestions are displayed as grayed-out ghost text and can be accepted with a single keystroke (e.g., Tab),
    which can be further modified by developers.
    Notably, our industry partner has deployed an in-use acceptance prediction method, called
    Circuit Breaker, which uses numerical features to predict accepted code suggestions, such as the
    number of preceding lines of code. We take the input features of this method as
    a reference when identifying candidate features.
    
    \subsubsection{Developer Feedback (abbreviated as Fdbk)} 
    Our industry partner provides us with some feedback records, which inspire us to design features. Our industry partner provides 25 feedback records between April 22--May 31, 2024,
    which include developers' experience and feedback towards the AI-assisted programming tool. 
    
    \subsubsection{AI-Developer Interaction Logs (abbreviated as Log)}
    Most importantly, our industry partner provides detailed developer-AI interaction logs within the IDE from developers with varying skill levels across multiple projects. 
    According to developer-AI interaction logs, we utilize available data to design features. 
    The main components of each log are detailed as follows:
    
    \begin{itemize}
    \item \textbf{Preceding Context:} refers to the content within the current file that exists before the generated code suggestions, serving as the contextual input for the code generation model.
    
    \item \textbf{Generated Code Suggestions:} refers to the code suggestions generated by the code generation model.
    
    \item \textbf{Max Generation Length:} refers to the maximum lines and characters of the generated code suggestions set by the developer.
    
    \item \textbf{Written Code:} refers to the code written between consecutive generation triggers within the same file.
    
    \item \textbf{Modified Code:} refers to any modifications applied to an accepted code suggestion before the next generation trigger.
     
    \item \textbf{IDE Type:} refers to the specific name of the IDE (e.g., VS Code).
    
    \item \textbf{IDE Version:} refers to the specific release number of the IDE, which defines the software environment where the code generation tool was operating.
    
    \item \textbf{Generation Time for Code Suggestions:} refers to the total duration from the moment a code suggestion is requested (e.g., by typing or pausing) to the moment the generated suggestions are fully displayed in the IDE.
    
    \item \textbf{Trigger Timestamp:} refer to the time point when the code generation function is triggered.
    
    \item \textbf{Accepted Length Ratio:} refers to the ratio of the accepted code length to the total length of the generated suggestion.
    
    \item \textbf{Acceptance Results:} refers to a boolean value indicating whether a generated code suggestion is accepted by the developer. 
    
    \end{itemize}
    
    \subsubsection{Literature References (abbreviated as Ref)} 
    To inform our feature design, we draw on prior work in several areas. This includes studies on pull request acceptance ~\cite{zhang2022pull, wu2023personalized}, research on programmer expectations and interaction with AI tools ~\cite{ciniselli2023source, ross2023programmer, erhabor2025measuring}, and analysis of how the coding environment impacts developer behavior ~\cite{ziegler2022productivity, zheng2023codegeex}.
    
    \subsection{Candidate Features}
    \label{candidate features}
    
    We design a compact feature set which spans three dimensions: developer habit, project preference, and in-situ context. Developer habit features and project preference features analyze historical characteristics of a developer or a project, respectively. In-situ code features capture a snapshot of the immediate circumstances surrounding a single code suggestion. 
    
    \subsubsection{Developer Habit Features} Developer habit features capture behavior patterns and preferences by analyzing a developer's recent coding history. To determine the optimal length of this history, we evaluate 7, 30, and 60-day windows and find that the shorter 7-day window identifies the same set of significant features. We adopt this shorter time window to improve computational efficiency.
    As shown in Table~\ref{tab:developer_habit_features},
    these features are further divided into three aspects, including developer code features,
    developer acceptance preference features and developer coding environment features.

    \textbf{Developer Code Features:} 
    
    \begin{itemize}
    \item \textbf{Preceding Context length.} 
    Our industry partner's Circuit Breaker uses context length to help predict acceptance. Inspired by this, we investigate how context length, measured in both lines and characters, may influence a developer's decision to accept a generated code suggestion. To investigate this, we characterize the length of a suggestion's preceding context by measuring the average number of lines (\textit{developer\_preceding\_lines}) and characters (\textit{developer\_preceding\_chars}) for a given developer.

    \item \textbf{Generated Code Length.}
    In preliminary developer feedback, a developer suggests a feature to automatically wrap single-line completions that exceed 120 characters, noting that code with few lines but a high character count may be rejected. 
    To investigate this, we measure the average number of lines in a developer's historically generated code suggestions (\textit{developer\_generated\_lines}) and  the average number of characters (\textit{developer\_generated\_chars}).
    
    \item \textbf{Modified Code length.}
    Both our preliminary developer feedback and literature review ~\cite{erhabor2025measuring} suggest that developers sometimes accept code suggestions
    and then manually revise code that require only minor corrections. 
    If developers make fewer modifications to the generated code, it may indicate a higher level of satisfaction with the generated content and a greater likelihood of accepting the subsequent code.
    To investigate this, we measure the average levenshtein distance   \cite{levenshtein1966binary} between an accepted suggestion and its final version (\textit{developer\_modified\_chars}), and the ratio of accepted suggestions that are subsequently modified (\textit{developer\_modified\_ratio}) for a given developer.
    
    \item \textbf{Written Code length.}
    The length of code a developer manually writes between suggestions may influence their subsequent acceptance decisions. To investigate this, we measure the average lines and characters of written code (\textit{developer\_written\_lines} and \textit{developer\_written\_chars}) for a given developer.
    
    \item \textbf{Generation Time.}
    In preliminary developer feedback,
    a developer mentions that slowness is the reason for rejecting code suggestions. 
    Accepted and rejected code suggestions may have different generation time. 
    To investigate this, we measure generation time for all presented, accepted, and rejected code suggestions for a given developer (\textit{developer\_presented\_time}, \textit{developer\_accepted\_time} and \textit{developer\_rejected\_time}).
    
    \item \textbf{Generation Interval.}
    In the developer feedback, several developers reject suggestions if they are triggered too frequently, as constant interruptions can break their concentration. 
    Therefore, we measure the average time elapsed between consecutive calls for code generation for a given developer (\textit{developer\_generated\_interval}).
    
    \item \textbf{Generation Counts.}
    Prior work has found that past developer engagement with intelligent tools predicts future adoption ~\cite{ciniselli2023source}. A rich interaction history may reflect a developer's confidence in the AI-assisted programming tool, which may increase the probability that they will accept future code suggestions. 
    We measure the total number of historically generated code suggestions (developer\_generated\_counts) for a given developer.
    
    \end{itemize}
    \begin{table}
        \caption{Overview of the features extracted to characterize developer Habits.}
        \label{tab:developer_habit_features}
        \resizebox{\textwidth}{!}{%
            \begin{tabular}{|@{}c|l|l|l@{}|}
                \hline
                \textbf{Dimension}                       &\textbf{Feature}                   & \textbf{Description} & \textbf{Source} \\
                \hline
                \multirow[c]{13}{*}{\textbf{Developer Code}}  & developer\_preceding\_lines                   & Average lines of preceding context in all logs for a developer                 & Cbk \\
                                                         & developer\_preceding\_chars                       & Average characters of preceding context in all logs for a developer                & Cbk \\
                                                         & developer\_generated\_lines                & Average lines of generated code suggestions in all logs for a developer                 & Fdbk \\
                                                         & developer\_generated\_chars                & Average characters of generated code suggestions in all logs for a developer                 & Fdbk \\
                                                         & developer\_modified\_chars                & Average characters of modified code suggestions in all logs for a developer                & Fdbk, Ref  \cite{erhabor2025measuring} \\
                                                         & developer\_modified\_ratio                & Average ratio of modified accepted code suggestions in all logs for a developer                & Fdbk, Ref  \cite{erhabor2025measuring} \\
                                                         & developer\_written\_lines                & Average lines of written code in all logs for a developer                 & Log \\
                                                         & developer\_written\_chars                & Average characters of written code in all logs for a developer                 & Log \\
                                                         & developer\_presented\_time                & Average generation-to-presentation time of code suggestions in all logs for a developer                 & Fdbk \\
                                                         & developer\_accepted\_time                & Average generation-to-presentation time of accepted code suggestions in all logs for a developer                 & Fdbk \\
                                                         & developer\_rejected\_time               & Average generation-to-presentation time of rejected code suggestions in all logs for a developer                 & Fdbk \\
                                                         & developer\_generated\_interval                & Average time interval between code generation triggers in all logs for a developer                 & Fdbk \\
                                                         & developer\_generated\_counts                & Total number of historically generated code suggestions in all logs for a developer               & Ref.  \cite{ciniselli2023source} \\

                \hline
                \multirow[c]{3}{*}{\textbf{Developer Acceptance Preference}}    & developer\_accepted\_length\_ratio & Average ratio of accepted length to total suggestion length in all logs for a developer.  & Ref  \cite{zhang2022pull} \\
                                                         & developer\_accepted\_ratio                & Average ratio of accepted to total generated code suggestions in all logs for a developer                  & Ref  \cite{zhang2022pull} \\
                                                         & developer\_accepted\_counts     & Total counts of accepted code suggestions in all logs for a developer       & Ref  \cite{zhang2022pull} \\
                \hline
                \multirow[c]{4}{*}{\textbf{Developer Coding Environment}} & developer\_IDE\_version\_counts         & Total counts of unique IDE versions in all logs for a developer                       & Log \\
                                                         & developer\_IDE\_type\_counts              & Total counts of unique IDE types in all logs for a developer                  & Log \\
                                                         & developer\_dominant\_IDE\_version   & The most frequent used IDE version in all logs for a developer & Log \\
                                                         & developer\_dominant\_IDE\_type    & The most frequent used IDE type in all logs for a developer  & Log \\
                \hline
            \end{tabular}
        }
    \end{table}
    
    \textbf{Developer Acceptance Preference Features:} 
    Prior work has found that a contributor's history of accepted pull requests is a strong predictor of future acceptance   \cite{zhang2022pull,wang2021accept}. Similarly, developer's past behavior with generated code suggestions may predict their future decisions. We measure the average accepted length ratio (\textit{developer\_accepted\_length\_ratio}); the ratio of accepted to total generated code suggestions (\textit{developer\_accepted\_ratio}); and the total number of accepted suggestions (\textit{developer\_accepted\_counts}).

    \textbf{Developer Coding Environment Features:} 
    The version of the underlying generative model may impact the quality of code suggestions and, consequently, developer acceptance. 
    While our industry partner does not provide direct model version data for confidentiality reasons, they have confirmed a positive correlation between IDE and model versions in our collected logs. We therefore use the IDE version as a practical proxy for model evolution when modeling acceptance. The most common IDE type is VS Code, and there are other IDE types inside the company.
    To characterize programming environment for a given developer, we measure the total number of unique IDE release versions and type names (\textit{developer\_IDE\_version\_counts} and \textit{developer\_IDE\_type\_counts}); the most frequent used IDE version and type (\textit{developer\_dominant\_IDE\_version} and \textit{developer\_dominant\_IDE\_type}).
    
    \subsubsection{Project Preference Features}
    Project preference features analyze historical characteristics of a project. 
    For example,
    developers may like to accept generated suggestions in early project stages, as these situations increase the need for coding guidance. Same to developer habit features, we also adopt a 7-day window when calculating project preference features. These features are also divided into three aspects, including project code features, project acceptance preference features and project programming environment features.
    
    \textbf{Project Code Features:} 
    
    \begin{itemize}
    \item \textbf{Preceding Comment Length.}
    The more comments indicate stricter programming standards  \cite{wu2023personalized}, which may influence the acceptance of generated code suggestions. We measure the average lines of comment per 1000 lines for a given project (\textit{project\_comment\_lines}).
    
    \item \textbf{Preceding Context Length.}
    Ross et al.  \cite{ross2023programmer} found that when project context is limited, developers may be in exploration mode and more likely to accept code suggestions. Analogous to developer preceding context features, we also measure the average lines and characters of preceding context for a given project (\textit{project\_preceding\_lines} and \textit{project\_preceding\_chars}). 
    
    \item \textbf{Generated Code Length.}
    Analogous to developer generated code features, we also measure the average lines and characters of historically generated code suggestions for a given project (\textit{project\_generated\_\allowbreak lines} and \textit{project\_generated\_chars}).
    
    \item \textbf{Generation Time.}
    Analogous to developer generation time features, we also measure the generation time for all presented, accepted, and rejected code suggestions for a given project (\textit{project\_presented\_time}, \textit{project\_accepted\_time} and \textit{project\_rejected\_time}).
    \end{itemize}
    
    \textbf{Project Acceptance Preference Features:} 
    Inspired by the finding that a project's history of accepting contributions is a strong predictor of its future behavior   \cite{zhang2022pull}, we introduce Project Acceptance Preference Features. Analogous to the developer acceptance features that capture individual habits, these project-level metrics quantify how well generated suggestions align with a project’s collective standards and needs. We measure the average accepted length ratio (\textit{project\_accepted\_\allowbreak length\_\allowbreak ratio}); the ratio of accepted to total generated code suggestions (\textit{project\_accepted\_\allowbreak ratio}); and the total number of accepted suggestions (\textit{project\_accepted\_\allowbreak counts}).

    \textbf{Project Coding Environment Features:} 
    Analogous to developer programming environment features, in order to describe project adaptation with the evolution of development environments, we measure the total number of unique IDE versions and types (\textit{project\_IDE\_version\_counts} and \textit{project\_IDE\_type\_counts}); the most frequent
    used IDE version and type (\textit{project\_dominant\_\allowbreak IDE\_\allowbreak version} and \textit{project\_dominant\_IDE\_\allowbreak type}).
    \begin{table}
        \caption{Overview of the features extracted to characterize Project Preferences.}
        \label{tab:project_preference_features}
        \resizebox{\textwidth}{!}{%
            \begin{tabular}{|@{}c|l|l|l@{}|}
                \hline
                \textbf{Dimension}                       &\textbf{Feature}                   & \textbf{Description} & \textbf{Source} \\
                \hline
                \multirow[c]{10}{*}{\textbf{Project Code Features}}  & project\_comment\_lines                   & Average lines of comment per 1000 lines in all logs for a project                 & Ref ~\cite{wu2023personalized} \\
                                                         & project\_preceding\_lines                       & Average lines of preceding context in all logs for a project                & Ref  \cite{ross2023programmer} \\
                                                         & project\_preceding\_chars                & Average characters of preceding context in all logs for a project                 & Ref \cite{ross2023programmer} \\
                                                         & project\_generated\_lines                & Average lines of generated code suggestions in all logs for a project                 & Fdbk \\
                                                         & project\_generated\_chars                & Average characters of generated code suggestions in all logs for a project                 & Fdbk \\
                                                         & project\_written\_lines                & Average lines of written code suggestions in all logs for a project                 & Log \\
                                                         & project\_written\_chars                & Average lines of written characters suggestions in all logs for a project                 & Log \\
                                                         & project\_presented\_time                & Average generation-to-presentation time of code suggestions in all logs for a project                 & Fdbk \\  
                                                         & project\_accepted\_time                & Average generation-to-presentation time of accepted code suggestions in all logs for a project                 & Fdbk \\ 
                                                         & project\_rejected\_time                & Average generation-to-presentation time of rejected code suggestions in all logs for a project                 & Fdbk \\ 
                \hline
                \multirow[c]{3}{*}{\textbf{Project Acceptance Preference}}    & project\_accepted\_length\_ratio & Average ratio of accepted length to total suggestion length in all logs for a project.  & Ref  \cite{zhang2022pull} \\
                                                         & project\_accepted\_ratio                & Average ratio of accepted to total generated code suggestions in all logs for a project                  & Ref \cite{zhang2022pull} \\
                                                         & project\_accepted\_counts     & Total counts of accepted code suggestions in all logs for a project       & Ref \cite{zhang2022pull} \\
                \hline
                \multirow[c]{4}{*}{\textbf{Project Coding Environment}} & project\_IDE\_version\_counts         & Total counts of unique IDE versions in all logs for a project                       & Log \\
                                                         & project\_IDE\_type\_counts              & Total counts of unique IDE types in all logs for a project                  & Log \\
                                                         & project\_dominant\_IDE\_version   & The most frequent used IDE version in all logs for a project & Log \\
                                                         & project\_dominant\_IDE\_type    & The most frequent used IDE type in all logs for a project  & Log \\
                                                         
                \hline
            \end{tabular}
        }
    \end{table}
    
    \subsubsection{In-situ Context Features}
    Unlike developer habit and project preference features that are aggregated from historical data, in-situ features capture a snapshot of the immediate circumstances surrounding a single generation event. They focus on the specific context of the generated code suggestion, the characteristics of the suggestion itself, and the coding environment at that precise moment. These features are designed from three aspects:

    \textbf{Preceding Context:} The preceding context is defined as all content within the file before the point of code generation. Inspired by our industry partner's Circuit Breaker, this set of features is extracted by analyzing this context, capturing metrics such as the number of lines (\textit{in-situ\_preceding\_lines}), number of characters (\textit{in-situ\_preceding\_chars} and \textit{in-situ\_preceding\_last\_line\_\allowbreak chars}), comment length (\textit{in-situ\_preceding\_comment\_chars}), and whether the context ends in an incomplete code block (\textit{in-situ\_preceding\_is\_incomplete}). The goal is to characterize the preceding content to understand its impact on the final acceptance outcome.
    
    \textbf{Generated Code Suggestion:} 
    Generated code suggestion itself may be of great importance. To investigate this, we measure the length of the generated code suggestion (\textit{in-situ\_generated\_lines} and \textit{in-situ\_generated\_chars}).
    Furthermore, we also consider the cosine similarity between generated code suggestion and preceding context (\textit{in-situ\_generated\_\allowbreak similarity}).
    
    \textbf{In-situ Coding Environment:} 
    Previous works  \cite{ziegler2022productivity,zheng2023codegeex} have suggested that developers are more willing to accept code outside working hours and have higher adoption rates when using certain IDE. 
    Furthermore,
    some developers set the max line or char limit of generated code suggestions.
    If the actual length of generated code suggestions is much different from developers' setting,
    they may have lower acceptance probability.
    In-situ coding environment may also influence on the acceptance of developer's decision. Therefore, we measure the generation time of in-situ code suggestion (\textit{in-situ\_generated\_time}); whether the generation timestamp is in work day (\textit{in-situ\_is\_workday}); the time period of the generated timestamp (\textit{in-situ\_time\_period}); the in-situ IDE (\textit{in-situ\_IDE\_type} and \textit{in-situ\_IDE\_version}); the max limit of generated code suggestion (\textit{in-situ\_max\_generated\_line} and \textit{in-situ\_max\_generated\_char}); the absolute delta between max generated length and actual one (\textit{in-situ\_abs\_lines\_delta} and \textit{in-situ\_abs\_chars\_delta}).

    \begin{table}
        \caption{Overview of In-situ Features.}
        \label{tab:insitu_features}
        \resizebox{\textwidth}{!}{%
            \begin{tabular}{|@{}c|l|l|l@{}|}
                \hline
                \textbf{Dimension}                       &\textbf{Feature}                   & \textbf{Description} & \textbf{Source} \\
                \hline
                \multirow[c]{5}{*}{\textbf{Preceding Context}}
                                                         & in-situ\_preceding\_lines                       & Average lines of preceding context of in-situ log                & Cbk \\
                                                         & in-situ\_preceding\_chars                & Aaverage characters of preceding context of in-situ log                 & Cbk \\
                                                         & in-situ\_preceding\_last\_line\_chars                & The characters of the last line of the preceding context of in-situ log                 & Cbk \\
                                                         & in-situ\_preceding\_comment\_chars                & The characters of comments of the preceding context of in-situ log                 & Cbk \\
                                                         & in-situ\_preceding\_is\_incomplete               & Whether the last line of the preceding context is an incomplete code block                 & Cbk \\
                \hline
                \multirow[c]{3}{*}{\textbf{Generated Code Suggestion}}    & in-situ\_generated\_lines & Lines of the generated code suggestion of in-situ log  & Fdbk \\
                                                         & in-situ\_generated\_chars                & Characters of the generated code suggestion of in-situ log                  & Fdbk \\
                                                         & in-situ\_generated\_similarity     & Cosine similarity between generated code and preceding context of in-situ log       &  Log \\
                \hline
                \multirow[c]{9}{*}{\textbf{In-situ Coding Environment}} & in-situ\_generation\_time         & Time to generate the code suggestion of in-situ log                       & Fdbk \\
                                                         & in-situ\_is\_workday              & Whether the generation time is a workday of in-situ log                  & Ref  \cite{ziegler2022productivity,zheng2023codegeex} \\
                                                         & in-situ\_time\_period   & The time period of the generation time of in-situ log (eg., 12:00--18:00) & Ref  \cite{ziegler2022productivity,zheng2023codegeex} \\
                                                         & in-situ\_IDE\_type    & The IDE type of in-situ log  & Log \\
                                                         & in-situ\_IDE\_version    & The IDE version of in-situ log  & Log \\
                                                         & in-situ\_max\_generated\_line    & Max line limit of generated code suggestions for in-situ log. & Log \\
                                                         & in-situ\_max\_generated\_char    & Max char limit of generated code suggestions for in-situ log. & Log \\
                                                         & in-situ\_abs\_lines\_delta    & Absolute delta between the developer-set maximum and actual generated lines. & Log \\
                                                         & in-situ\_abs\_chars\_delta    & Absolute delta between the developer-set maximum and actual generated characters. & Log \\
                                                         
                \hline
            \end{tabular}
        }
    \end{table}

    \subsection{Significance Analysis}
    In the previous subsection, we design multi-dimensional features. However, not all of these features can effectively distinguish between accepted and rejected code suggestions. We therefore employ significance testing to isolate the most influential ones—those showing a significant statistical difference between accepted and rejected code suggestions.
    
    \subsubsection{Methodology}
    We perform statistical analyses to identify the significant features of accepted code suggestions compared with rejected ones, which consists of three steps:
    
    \textbf{Step 1: Dataset Construction.} We construct a dataset from the developer-AI interaction logs provided by our industry partner, including 66,239 entries from 113 developers across 12 projects collected over six months. Then, we extract candidate features as illustrated in Section ~\ref{candidate features}.
    
    \textbf{Step 2: Testing Statistical Significance.} 
    According to previous work  \cite{khatoonabadi2023wasted},
    we apply the Mann-Whitney U test with a 95\% confidence level (i.e., $\alpha = 0.05$)~  \cite{mann1947test},
    so as to test the statistical difference between the features of accepted and rejected code suggestions. We use this non-parametric test because we cannot assume a normal distribution for our features. A feature is considered statistically significant if its p-value, adjusted for multiple comparisons, is less than 0.05.
    
    \textbf{Step 3: Testing Practical Significance.} While statistical significance verifies whether a difference exists, we also test for practical significance using Cliff's delta  \cite{khatoonabadi2023wasted} to estimate the magnitude of the difference (i.e., effect size). The value of Cliff's delta ($d$) ranges from -1 to +1; a positive value indicates that the feature's values are typically greater for accepted suggestions than for rejected ones, while a negative value implies the opposite. This directional difference is noted as "Greater" or "Less" in the significance tables. For instance, a "Greater" result for the \textit{developer\_accepted\_ratio} feature indicates that accepted suggestions are characterized by a higher average ratio than rejected ones. Following prior work ~\cite{khatoonabadi2023wasted}, we consider a feature to have practical significance if $|d| > 0.147$. Finally, a feature is considered significant if it is both statistically and practically significant.
    
    \subsubsection{Findings}

     \begin{table}
        \caption{Overview of developer Habit Feature Significance}
        \label{tab:overview of developer habit feature significance}
        \scriptsize
        \centering
        \setlength{\tabcolsep}{4pt}
        \renewcommand{\arraystretch}{1.1}
        \begin{adjustbox}{width=\linewidth}
            \begin{tabular}{|c|p{0.26\linewidth}|p{0.25\linewidth}|c|}
                \hline 
                \textbf{Significance} & \textbf{Feature Category} & \textbf{Feature Name} & \textbf{Accepted vs. Rejected} \\
                \hline
                \multirow[c]{8}{*}{\textbf{Significant}} & \multirow[c]{5}{*}{\textbf{Developer Code}} & developer\_generated\_interval & Greater \\
                                                         & & developer\_generated\_lines & Greater \\
                                                         & & developer\_generated\_chars & Greater \\
                                                         & & developer\_presented\_time & Greater \\
                                                         & & developer\_accepted\_time & Greater \\
                \cline{2-4}
                                                         & \multirow[c]{3}{*}{\textbf{Developer Acceptance Preference}} & developer\_accepted\_length\_ratio & Greater \\
                                                         & & developer\_accepted\_ratio & Greater \\
                                                         & & developer\_accepted\_counts & Greater \\
                \hline
                \multirow[c]{12}{*}{\textbf{Insignificant}} & \multirow[c]{8}{*}{\textbf{Developer Code}} & developer\_preceding\_lines & -- \\
                                                         & & developer\_preceding\_chars & -- \\
                                                         & & developer\_written\_lines & -- \\
                                                         & & developer\_written\_chars & -- \\
                                                         & & developer\_modified\_chars & -- \\
                                                         & & developer\_modified\_ratio & -- \\
                                                         & & developer\_rejected\_time & -- \\
                                                         & & developer\_generated\_counts & -- \\
                \cline{2-4}
                                                         & \multirow[c]{4}{*}{\textbf{Developer Coding Environment}} & developer\_IDE\_version\_counts & -- \\
                                                         & & developer\_IDE\_type\_counts & -- \\
                                                         & & developer\_dominant\_IDE\_version & -- \\
                                                         & & developer\_dominant\_IDE\_type & -- \\
                \hline
            \end{tabular}
        \end{adjustbox}
    \end{table}
    
    Our analysis reveals that features related to developer habits and project preferences are the most significant features in code suggestion acceptance. 
    In contrast, many in-situ features show limited statistical significance in our analysis.  
    In the following, we discuss the significance of each feature dimension in more detail, with summaries provided in Tables~\ref{tab:overview of developer habit feature significance}, \ref{tab:project_prefs_overview}, and \ref{tab:insitu_feature_significance}.
    
    \textbf{Accepted suggestions usually have a more positive acceptance history than rejected ones.} At the developer level, accepted suggestions are characterized by a higher developer acceptance rate (\textit{developer\_accepted\_ratio}) and a greater number of previously accepted suggestions (\textit{developer\_accepted\_counts}), as shown in Table~\ref{tab:overview of developer habit feature significance}. This suggests a positive feedback loop: satisfaction with the tool leads to continued engagement and trust. Similarly, at the project level, a project's collective acceptance history is significantly important (Table~\ref{tab:project_prefs_overview}). Accepted suggestions are typically found in projects where previous suggestions are also frequently accepted, as indicated by higher \textit{project\_accepted\_ratio} and \textit{project\_accepted\_counts}. This points to a strong alignment between the AI-assisted programming tool's outputs and the project’s specific requirements.
    
    \textbf{Accepted code suggestions usually involve longer wait times than rejected ones.} Interestingly, 
    accepted code suggestions have
    longer generation times (\textit{developer\_presented\_\allowbreak time} and \textit{developer\_accepted\_time}). While counterintuitive from a user experience perspective, this could indicate that the model produces higher-quality, more relevant code when given more time for inference, which developers are ultimately more willing to accept despite the delay.

    \textbf{Accepted code suggestions usually follow longer intervals between triggers.} We found that accepted suggestions are characterized by a longer time interval between suggestion triggers (\textit{developer\_generated\_interval}). This may be because frequent, unsolicited suggestions can disrupt a developer's concentration and workflow. Longer pauses between suggestions may reduce this interference, allowing developers to more thoughtfully consider and accept the code proposed by the model.

    \begin{table}
        \caption{Overview of Project Preference Feature Significance}
        \label{tab:project_prefs_overview}
        \scriptsize
        \centering
        \setlength{\tabcolsep}{4pt}
        \renewcommand{\arraystretch}{1.1}
        \begin{adjustbox}{width=\linewidth}
            \begin{tabular}{|c|p{0.26\linewidth}|p{0.25\linewidth}|c|}
                \hline
                \textbf{Significance} & \textbf{Feature Category} & \textbf{Feature Name} & \textbf{Accepted vs. Rejected} \\
                \hline
                \multirow[c]{7}{*}{\textbf{Significant}} & \multirow[c]{4}{*}{\textbf{Project Code}} & project\_preceding\_lines & Less \\
                                                         & & project\_preceding\_chars & Less \\
                                                         & & project\_presented\_time & Greater \\
                                                         & & project\_accepted\_time & Greater \\
                \cline{2-4}
                                                         & \multirow[c]{3}{*}{\textbf{Project Acceptance Preference}} & project\_accepted\_length\_ratio & Greater \\
                                                         & & project\_accepted\_ratio & Greater \\
                                                         & & project\_accepted\_counts & Greater \\
                \hline
                \multirow[c]{10}{*}{\textbf{Insignificant}} & \multirow[c]{6}{*}{\textbf{Project Code}} & project\_comment\_lines & -- \\
                                                         & & project\_generated\_lines & -- \\
                                                         & & project\_generated\_chars & -- \\
                                                         & & project\_written\_lines & -- \\
                                                         & & project\_written\_chars & -- \\
                                                         & & project\_rejected\_time & -- \\
                \cline{2-4}
                                                         & \multirow[c]{4}{*}{\textbf{Project Coding Environment}} & project\_IDE\_version\_counts & -- \\
                                                         & & project\_IDE\_type\_counts & -- \\
                                                         & & project\_dominant\_IDE\_version & -- \\
                                                         & & project\_dominant\_IDE\_type & -- \\
                \hline
            \end{tabular}
        \end{adjustbox}
    \end{table}
    
    \textbf{Accepted code suggestions usually have smaller preceding contexts at the project level than rejected ones.} 
    In comparison to rejected code suggestions,
    accepted suggestions have the smaller (\textit{project\_preceding\_lines} and \textit{project\_preceding\_chars}) preceding context in the project. 
    This scenario may occur in the early stages of a new feature or file (e.g., initializing a class, setting up a module). In these situations, developers may be more open to assistance to bootstrap their work or explore implementation ideas.
    
    \textbf{In-situ features are nuanced and hard to measure effectively.} While many in-situ features appear statistically insignificant (Table~\ref{tab:insitu_feature_significance}), this does not diminish their real-world importance. For instance, the similarity between a suggestion and the preceding context (\textit{in-situ\_generated\_similarity}) is not significant. This may be because developers often seek suggestions that are causally related to the preceding code (i.e., what should come \emph{next}), not merely similar to it. This causal relationship is semantically complex and difficult to measure with simple metrics like cosine similarity. The most critical factor—the match between the generated code and the developer’s ultimate intent—is unobservable at the moment of generation, making it challenging to identify strong, predictive in-situ signals from logs alone. Indeed, prior work ~  \cite{liang2024large,tan2024far,ciniselli2023source} confirms that one of the strongest predictors of acceptance is the semantic similarity between the generated suggestion and the one the developer eventually writes, a metric that is, by definition, unavailable at prediction time.
    
    \textbf{Impact of Model and IDE Version Updates on Acceptance.} The only significant in-situ feature is the IDE version (\textit{in-situ\_IDE\_version}),
    with accepted code suggestions having older versions. As IDE version serves as a proxy for the underlying model version in our deployment, this pattern likely reflects model-version effects. 
    Our industry partner explains that as the IDE and its associated libraries are updated over time, the model, trained on older data, may generate less effective or relevant code, leading to a poorer user experience and lower acceptance rates in newer IDE versions. This underscores the critical need to account for changes in model versions over time when analyzing and predicting acceptance.
    
    \begin{table}
        \caption{Overview of In-situ Feature Significance}
        \label{tab:insitu_feature_significance}
        \scriptsize
        \centering
        \setlength{\tabcolsep}{4pt}
        \renewcommand{\arraystretch}{1.1}
        \begin{adjustbox}{width=\linewidth}
            \begin{tabular}{|c|p{0.26\linewidth}|p{0.25\linewidth}|c|}
                \hline
                \textbf{Significance} & \textbf{Feature Category} & \textbf{Feature Name} & \textbf{Accepted vs. Rejected} \\
                \hline
                \textbf{Significant} & \textbf{In-situ Coding Environment} & in-situ\_IDE\_version & Less \\
                \hline
                \multirow[c]{16}{*}{\textbf{Insignificant}} & \multirow[c]{5}{*}{\textbf{Preceding Context}} & in-situ\_preceding\_lines & -- \\
                                                         & & in-situ\_preceding\_chars & -- \\
                                                         & & in-situ\_preceding\_last\_line\_chars & -- \\
                                                         & & in-situ\_preceding\_comment\_chars & -- \\
                                                         & & in-situ\_preceding\_is\_incomplete & -- \\
                \cline{2-4}
                                                         & \multirow[c]{3}{*}{\textbf{Generated Code Suggestion}} & in-situ\_generated\_lines & -- \\
                                                         & & in-situ\_generated\_chars & -- \\
                                                         & & in-situ\_generated\_similarity & -- \\
                \cline{2-4}
                                                         & \multirow[c]{8}{*}{\textbf{In-situ Coding Environment}} & in-situ\_generation\_time & -- \\
                                                         & & in-situ\_is\_workday & -- \\
                                                         & & in-situ\_time\_period & -- \\
                                                         & & in-situ\_IDE\_type & -- \\
                                                         & & in-situ\_IDE\_version & -- \\
                                                         & & in-situ\_max\_generated\_line & -- \\
                                                         & & in-situ\_abs\_lines\_delta & -- \\
                                                         & & in-situ\_abs\_chars\_delta & -- \\
                \hline
            \end{tabular}
        \end{adjustbox}
    \end{table}
    
    \bigskip
    \begin{tcolorbox}
        \noindent\textbf{\emph{Answer to RQ\textsubscript{1}.}}  
       We find significant differences related to developer habits and project preferences between accepted and rejected code suggestions. At the developer level, accepted suggestions are characterized by the developer's significantly higher historical acceptance count and ratio, and longer generation intervals. At the project level, they are characterized by the project's significantly higher historical acceptance count and ratio, and a shorter preceding code context. In contrast, most in-situ features show limited statistical significance, likely because the most critical factor—alignment with a developer's true intent—is unobservable at generation time. However, we find that accepted code suggestions having older IDE versions, suggesting the model has become less adapted to evolving coding corpora, which may degrade the user experience over time.
    \end{tcolorbox}
    \medskip
    
\section{RQ2: Code Suggestion Acceptance Prediction}
\label{sec:rq2}

Building on the significant features identified in RQ\textsubscript{1} across developer habits, project preferences, and in-situ context, RQ\textsubscript{2} turns these signals into a prediction task. We first distill a compact predictive set via correlation analysis, then design \emph{CSAP} to integrate these features, and finally evaluate it against strong baselines with feature contribution analysis.

\subsection{Correlation Analysis}
\subsubsection{Feature Correlation Analysis}
Significant features may include highly correlated ones, and directly inputting highly correlated features into models may bias results. We therefore compute Spearman rank correlations between feature pairs ~\cite{khatoonabadi2023wasted}. Under the premise of p-value < 0.05 (statistical significance), we treat \(|\rho| > 0.8\) as strong correlation and keep a single representative feature per correlated cluster to reduce redundancy.

Our analysis revealed: (1) among developer habit features, only \textit{developer\_accepted\_length\_ratio} and \textit{developer\_accepted\_ratio} has \(|\rho|>0.8\); we retain \textit{developer\_accepted\_ratio}. (2) Among project preference features, \textit{project\_preceding\_lines} and  \textit{project\_preceding\_chars} has \(|\rho|>0.8\); we retain \textit{project\_preceding\_lines}. (3) \textit{project\_accepted\_time}, \textit{project\_accepted\_length\_ratio}, and \textit{project\_accepted\allowbreak \_ratio} are mutually correlated (\(|\rho|>0.8\)); we retain \textit{project\_accepted.\_ratio}. All other features without high correlation are kept.

\subsubsection{Independent Features for Prediction}
\label{sec:independent_features}
After applying significance analysis and correlation pruning, we derive the final feature set for prediction. As detailed in Table~\ref{tab:independent_features}, this set comprises features from three key categories: developer habit, project preference and in-situ context.

\begin{table}
    \caption{Independent Features Used for Prediction}
    \label{tab:independent_features}
    \scriptsize
    \centering
    \setlength{\tabcolsep}{4pt}
    \renewcommand{\arraystretch}{1.1}
        \begin{tabular}{|c|l|}
            \hline
            \textbf{Feature Category} & \textbf{Feature Name} \\
            \hline
            \multirow[c]{7}{*}{\textbf{Developer Habit Features}} & developer\_generated\_interval \\
                                                                 & developer\_generated\_lines \\
                                                                 & developer\_generated\_chars \\
                                                                 & developer\_presented\_time \\
                                                                 & developer\_accepted\_time \\
                                                                 & developer\_accepted\_ratio \\
                                                                 & developer\_accepted\_counts \\
            \hline
            \multirow[c]{4}{*}{\textbf{Project Preference Features}} & project\_preceding\_lines \\
                                                                 & project\_presented\_time \\
                                                                 & project\_accepted\_ratio \\
                                                                 & project\_accepted\_counts \\
            \hline
            \textbf{In-situ context Features} & in-situ\_IDE\_version \\
            \hline
        \end{tabular}
\end{table}

\subsection{Method Design}
\label{subsec:method_design}

\begin{figure}[!htbp]
	\centering
	\includegraphics[trim = {0 2.5cm 0 0}, clip,width=0.80\textwidth]{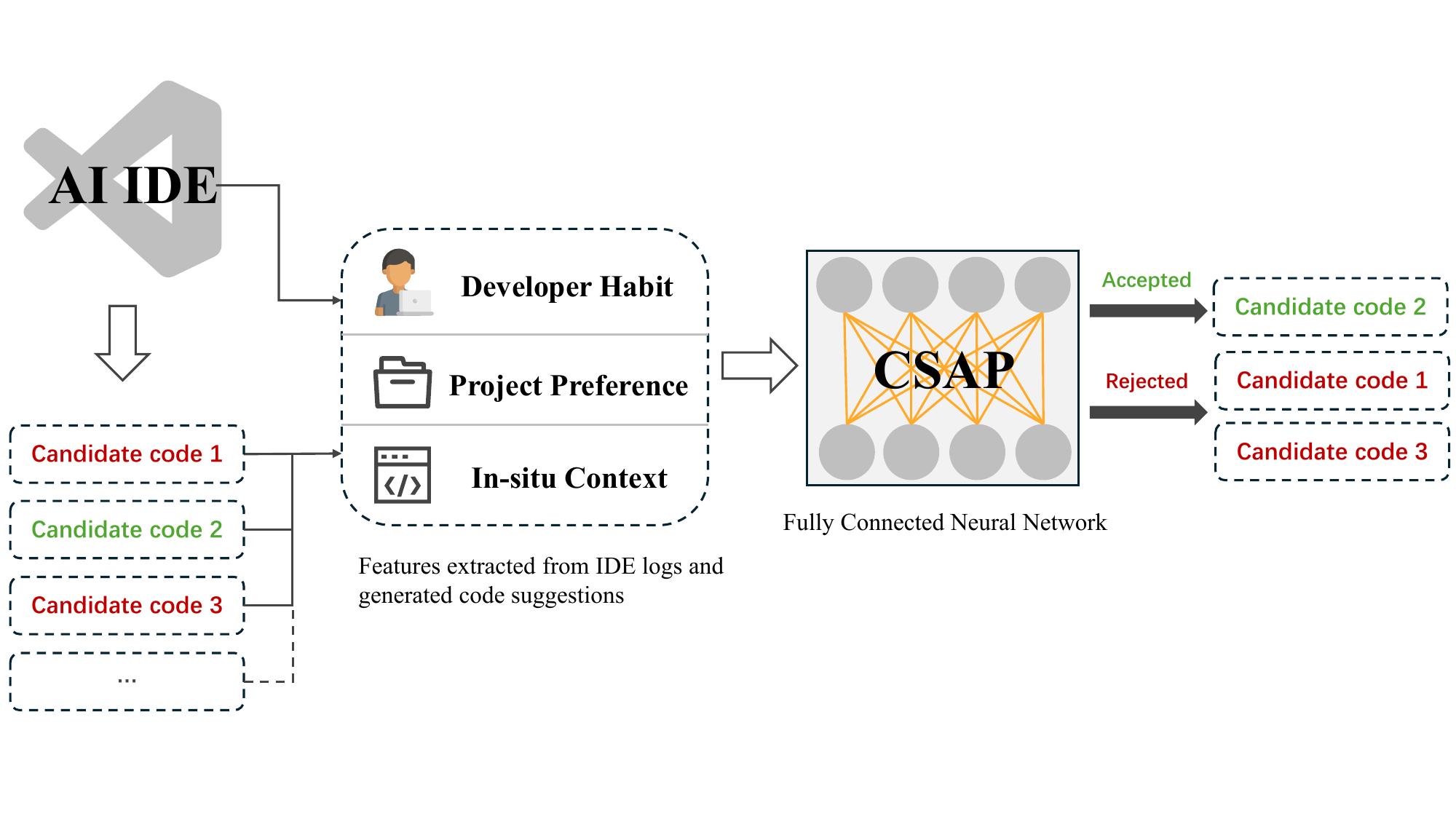}
	\caption{Overview of our designed method \emph{CSAP}}	\label{fig:pcap_overview}
\end{figure}

We propose \emph{CSAP} (Code Suggestion Acceptance Prediction), which formulates acceptance prediction as binary classification in Figure \ref{fig:pcap_overview}. \emph{CSAP} combines three sources: developer habit, project preference and in-situ context characteristics. 
In the training phase, the input features are first normalized and concatenated to form an \(M\)-dimensional vector, where \(M\) denotes the total number of features. This vector is then fed into a simple fully connected neural network, which uses a sigmoid function to map it to a probability.
If the probability is larger or equal to threshold,
the generated code suggestion is predicted as accept,
otherwise it is predicted as rejected.
We discuss the parameter setting in the subsection \ref{sub_parameter}.
In the testing phase,
we extract the same features and predict acceptance with the trained model.

Previous work on GitHub Copilot found that developers rejected approximately 70\% of the suggestions shown to them \cite{ziegler2022productivity}.
In the real developer-AI interaction logs provided by our industry partner,
the majority of code suggestions are also rejected.
Due to the imbalanced distribution of samples in the dataset, 
\emph{CSAP} employs the class-balanced binary cross-entropy as the loss function and utilizes the AdamW optimizer ~\cite{loshchilov2019decoupled} for gradient descent to train the predictive model.
More specifically,
Equations \ref{eq:weighted_loss} defines the class-balanced binary cross-entropy. Here, \(y_i \in \{0,1\}\) is the ground-truth label (1=accept, 0=reject), \(\hat{y}_i\) is the model score before the sigmoid, and \(\sigma(\cdot)\) maps the score to a probability.
The weights \(w_{+}\) and \(w_{-}\) are the proportion of negative samples and positive samples,
which are set inversely to class frequencies so that the minority class contributes more to the loss, mitigating the accept/reject imbalance. When accepted samples are rarer, \(w_{+}\) becomes larger, increasing the penalty for missing accepted cases and improving probability calibration.

\begin{equation}
\label{eq:weighted_loss}
L_{weight} = -\frac{1}{N} \sum_{i=1}^{N} \Bigl[ w_{+}\, y_i \, \log\!\bigl(\sigma(\hat{y}_i)\bigr)
+ w_{-}\, (1 - y_i) \, \log\!\bigl(1 - \sigma(\hat{y}_i)\bigr) \Bigr]
\end{equation}

\subsection{Experimental Setup}
\label{sec:experimental_setup}

\subsubsection{Dataset Construction}

To experimentally evaluate the \emph{CSAP} prediction method, we construct an experimental dataset based on the code acceptance characteristic dataset containing 66,239 developer-AI interaction logs. We sort the dataset by time from earliest to latest, taking the first 80\% of data as the training set and the last 20\% as the test set.
The training set has 52,991 entries, which is used to train the model for acceptance prediction.
The test set contains 13,248 entries, which is called the "imbalanced test set" in this paper.

To better validate the method's effectiveness, we construct a "balanced test set" by retaining all accepted instances from the imbalanced test set and subsampling an equal number of rejected instances.

\subsubsection{Evaluation Metrics}
\label{sec:evaluation_metrics}
According to previous work  \cite{Fan2018Early}, we evaluate with Accuracy, Precision, Recall, F1-score. Accuracy is the fraction of predictions that match the true accept/reject label. Precision (accept) is the share of truly accepted items among those predicted as accepted, and Recall (accept) is the share of truly accepted items that we correctly predict as accepted; the F1-score (accept) is the harmonic mean of these two. We report Precision/Recall/F1 for the reject class analogously. 

We also evaluate results by Binary Cross-Entropy (BCE), which is suggested by our industry partner.
BCE measures the discrepancy between predicted probabilities and actual labels, thereby capturing the quality of the predictions and penalizing overconfident mistakes.
The computation of Binary Cross-Entropy is as follows:
\begin{equation}
\label{eq:cross_entropy}
L(y, \hat{y}) = -\frac{1}{N} \sum_{i=1}^{N} \left[ y_i \log(\hat{y_i}) + (1-y_i) \log(1-\hat{y_i}) \right]
\end{equation}

Here, $\hat{y}$ is the vector of predicted probabilities for all data, $y$ is the vector of labels for all data, $N$ is the number of data items, $y_i$ is the true label of the $i$-th sample, and $\hat{y}_i$ is the predicted probability that the $i$-th sample is accepted.
For methods using LLM that output a definitive 'accept'/'reject' decision (a hard label) instead of a probability, we convert these outcomes to 0.999999 and 0.000001, respectively. This smoothing avoids \(\log(0)\), approximates deterministic predictions as near-certain probabilities, and keeps BCE finite and comparable across methods.
Values of Accuracy, Precision, Recall, F1-score are between 0 and 1, and larger values mean better performance.
In contrast, larger Binary Cross-Entropy mean worse performance.

\subsubsection{Parameter Settings}
\label{sub_parameter}

We sweep the decision threshold from 0.1 to 0.9 (Table~\ref{tab:threshold_pivot}). On the imbalanced test set, most metrics reach their best values to the right of 0.5 ($\geq 0.6$), and when a metric peaks elsewhere, 0.5 is within a small margin. On the balanced set, all metrics peak to the left of 0.5 ($\approx 0.4$). Since no single threshold dominates across both sets and 0.5 is consistently near-optimal, we retain 0.5 as a stable, distribution-agnostic decision threshold.

\begin{table}[htbp]
\footnotesize
\caption{\emph{CSAP} Performance Metrics with Thresholds as Columns.}
\centering
\setlength{\tabcolsep}{4pt}
\renewcommand{\arraystretch}{1.1}
\begin{tabular}{|l|l|*{9}{c|}}
\hline
\multicolumn{2}{|c|}{\textbf{Metric}} & \multicolumn{9}{c|}{\textbf{Probability Threshold}} \\ \cline{3-11}
\multicolumn{2}{|c|}{} & \textbf{0.1} & \textbf{0.2} & \textbf{0.3} & \textbf{0.4} & \textbf{0.5} & \textbf{0.6} & \textbf{0.7} & \textbf{0.8} & \textbf{0.9} \\ \hline

\multirow{4}{*}{\textbf{Imbalanced}}
& Accuracy & 0.956 & 0.965 & 0.970 & 0.972 & 0.973 & 0.974 & 0.975 & 0.975 & 0.975 \\ \cline{2-11}
& Accept F1 & 0.793 & 0.828 & 0.848 & 0.857 & 0.861 & 0.864 & 0.866 & 0.864 & 0.862 \\ \cline{2-11}
& Reject F1 & 0.976 & 0.981 & 0.984 & 0.985 & 0.985 & 0.986 & 0.986 & 0.986 & 0.986 \\ \cline{2-11}
& Cross-Entropy & 0.604 & 0.478 & 0.410 & 0.381 & 0.368 & 0.358 & 0.349 & 0.351 & 0.351 \\ \hline

\multirow{4}{*}{\textbf{Balanced}}
& Accuracy & 0.917 & 0.920 & 0.922 & 0.924 & 0.922 & 0.921 & 0.921 & 0.916 & 0.910 \\ \cline{2-11}
& Accept F1 & 0.914 & 0.916 & 0.918 & 0.919 & 0.918 & 0.916 & 0.916 & 0.910 & 0.902 \\ \cline{2-11}
& Reject F1 & 0.921 & 0.923 & 0.926 & 0.928 & 0.927 & 0.925 & 0.926 & 0.921 & 0.916 \\ \cline{2-11}
& Cross-Entropy & 1.142 & 1.109 & 1.071 & 1.049 & 1.071 & 1.093 & 1.093 & 1.159 & 1.246 \\ \hline
\end{tabular}
\label{tab:threshold_pivot}
\end{table}

\subsection{Experimental Results}
\subsubsection{Prediction Performance}
\label{subsubsec:prediction_performance}
In order to evaluate the performance of \emph{CSAP}, we compare \emph{CSAP} against several baseline methods:

\textbf{(1) Circuit Breaker Model.} The model's approach is to predict the acceptance of code suggestions based on numerical features, such as the number of preceding lines of code. This method is a practical code acceptance prediction technique used internally by our industry partner.

\textbf{(2) Direct LLM Call (Qwen2.5-Coder-32B).} With the rise of powerful Large Language Models (LLMs), an intuitive baseline is to prompt an LLM with the same features used by \emph{CSAP} to predict code acceptance directly. To structure the input, we designed a four-part prompt: (i) instructions telling the model to predict code acceptance, (ii) definitions of unique start/end markers for each feature, (iii) a rule that limits the output to either "accept" or "reject", and (iv) the actual feature values wrapped in their markers. This method treats the task as a general, in-context learning problem. We use it as a strong baseline to demonstrate the effectiveness of our specialized \emph{CSAP} model, which is designed to handle structured numerical features. For this baseline, we use \textit{Qwen2.5-Coder-32B}, a model deployed in production by our industry partner.

\begin{table}[htbp]
\footnotesize
\caption{CSAP and Baseline Performance on Imbalanced and Balanced Test Sets.}
\centering
\setlength{\tabcolsep}{4pt}
\renewcommand{\arraystretch}{1.2}
\begin{adjustbox}{width=\linewidth}
\begin{tabular}{|c|c|c|ccc|ccc|c|}
\hline
\multirow{2}{*}{\textbf{Dataset}} &
\multirow{2}{*}{\textbf{Method}} &
\multirow{2}{*}{\textbf{Acc.}} &
\multicolumn{3}{c|}{\textbf{Accept}} &
\multicolumn{3}{c|}{\textbf{Reject}} &
\multirow{2}{*}{\makecell{\textbf{Cross-}\\\textbf{Entropy}}}
\\ \cline{4-9}
&&& \textbf{Prec.} & \textbf{Rec.} & \textbf{F1} & \textbf{Prec.} & \textbf{Rec.} & \textbf{F1} & \\ \hline

\multirow{4}{*}{Imbalanced}
& CSAP            & \textbf{0.973} & \textbf{0.858} & \textbf{0.864} & \textbf{0.861} & \textbf{0.986} & 0.985           & \textbf{0.985} & \textbf{0.368} \\ \cline{2-10}
& Direct LLM Call & 0.864          & 0.075          & 0.038          & 0.051          & 0.904          & 0.951           & 0.927          & 1.881 \\ \cline{2-10}
& Circuit Breaker & 0.574          & 0.041          & 0.153          & 0.064          & 0.874          & 0.618           & 0.724          & 5.891 \\ \hline

\multirow{4}{*}{Balanced}
& CSAP            & \textbf{0.922} & \textbf{0.977} & \textbf{0.864} & \textbf{0.917} & \textbf{0.878} & \textbf{0.979} & \textbf{0.926} & \textbf{1.082} \\ \cline{2-10}
& Direct LLM Call & 0.493          & 0.421          & 0.038          & 0.070          & 0.496          & 0.948           & 0.651          & 7.006 \\ \cline{2-10}
& Circuit Breaker & 0.384          & 0.284          & 0.153          & 0.199          & 0.420          & 0.614           & 0.499          & 8.514 \\ \hline
\end{tabular}
\end{adjustbox}
\label{tab:pcap_combined_results}
\end{table}

Table~\ref{tab:pcap_combined_results} summarizes \emph{CSAP} and baseline results on the metrics in Section~\ref{sec:evaluation_metrics}.
On the imbalanced test set, CSAP reaches 0.973 accuracy; for accepted code, precision/recall/F1 are 0.858/0.864/0.861, respectively; for rejected code, precision/recall/F1 are 0.986/0.985/0.985, respectively; BCE is 0.368.

Compared to the Circuit Breaker model and the direct LLM call,
 \emph{CSAP} improves the accuracy by 12.6\% and 69.5\% on imbalanced dataset,
 and improves the accuracy by 87.0\% and 140.1\% on balanced dataset.
 Across both imbalanced and balanced sets, \emph{CSAP} consistently outperforms the Circuit Breaker model and the direct LLM call on all metrics. 
Both baseline models perform poorly, particularly on the balanced test set, demonstrating their inability to handle class imbalance and effectively predict code acceptance. In contrast, \emph{CSAP} delivers the highest accuracy and lowest cross-entropy, indicating superior calibration and discrimination for accepted code.

\subsubsection{Feature Contribution Analysis}

\begin{table}[htbp]
\footnotesize
\caption{CSAP Performance with Different Feature Sets on Imbalanced and Balanced Test Sets}
\centering
\setlength{\tabcolsep}{4pt}
\renewcommand{\arraystretch}{1.2}
\begin{adjustbox}{width=\linewidth}
\begin{tabular}{|c|c|c|ccc|ccc|c|}
\hline
\multirow{2}{*}{\textbf{Dataset}} &
\multirow{2}{*}{\textbf{Method}} &
\multirow{2}{*}{\textbf{Acc.}} &
\multicolumn{3}{c|}{\textbf{Accept}} &
\multicolumn{3}{c|}{\textbf{Reject}} &
\multirow{2}{*}{\makecell{\textbf{Cross-Entropy}}}
\\ \cline{4-9}
&&& \textbf{Prec.} & \textbf{Rec.} & \textbf{F1} & \textbf{Prec.} & \textbf{Rec.} & \textbf{F1} & \\ \hline

\multirow{5}{*}{Imbalanced}
& All Features            & \textbf{0.973} & \textbf{0.858} & \textbf{0.864} & \textbf{0.861} & \textbf{0.986} & \textbf{0.985} & \textbf{0.985} & \textbf{0.368} \\ \cline{2-10}
& w/o In-situ Context        & 0.899          & 0.480          & 0.739          & 0.582          & 0.971          & 0.916          & 0.942          & 1.397 \\ \cline{2-10}
& w/o Developer Habits         & 0.903          & 0.494          & 0.786          & 0.607          & 0.976          & 0.915          & 0.945          & 1.342 \\ \cline{2-10}
& w/o Project Preferences & 0.965          & 0.806          & 0.842          & 0.822          & 0.983          & 0.978          & 0.981          & 0.480 \\ \cline{2-10} 
\hline

\multirow{5}{*}{Balanced}
& All Features            & \textbf{0.922} & \textbf{0.977} & \textbf{0.864} & \textbf{0.917} & \textbf{0.878} & \textbf{0.979} & \textbf{0.926} & \textbf{1.082} \\ \cline{2-10}
& w/o In-situ Context        & 0.826          & 0.895          & 0.739          & 0.810          & 0.778          & 0.914          & 0.840          & 2.399 \\ \cline{2-10}
& w/o Developer Habits         & 0.853          & 0.908          & 0.786          & 0.842          & 0.811          & 0.920          & 0.862          & 2.033 \\ \cline{2-10}
& w/o Project Preferences & 0.909          & 0.973          & 0.842          & 0.903          & 0.861          & 0.977          & 0.915          & 1.251 \\ \cline{2-10}
\hline
\end{tabular}
\end{adjustbox}
\label{tab:feature_combined}
\end{table}

To evaluate whether developer habit features, project preference features, and in-situ context features all contribute to \emph{CSAP}'s prediction performance, we conducted three experiments under the same experimental settings as Section~\ref{sec:experimental_setup}, each removing one aspect of features. We compared the results of these three experiments with the complete \emph{CSAP} method to analyze the effectiveness of the three feature aspects in improving prediction performance. The experimental results are shown in Table~\ref{tab:feature_combined}. 

The results show that after removing developer habit features, project preference features, and in-situ context features respectively, the model's performance decreases on accuracy, precision, recall, and F1-score for both accepted and rejected code suggestion, and increases on binary cross-entropy, indicating that all proposed features contribute to improving the model's prediction performance. Among these, removing in-situ context features results in the most significant degradation, highlighting their impact; developer habit features also matter, while project preferences have moderate effect.

\begin{table}[htbp]
\scriptsize
\caption{Individual feature ablation on Imbalanced and Balanced test sets ($\Delta$ = All Features $-$ Model w/o feature).}
\centering
\setlength{\tabcolsep}{4pt} 
\renewcommand{\arraystretch}{1.1}
\begin{adjustbox}{width=\linewidth}
\begin{tabular}{|l|c|c|c|c|}
\hline
\multirow{2}{*}{\textbf{Feature Name}} & \multicolumn{2}{c|}{\textbf{Imbalanced}} & \multicolumn{2}{c|}{\textbf{Balanced}} \\ \cline{2-5} 
 & \textbf{Accuracy} & \makecell{\textbf{Cross-Entropy}} & \textbf{Accuracy} & \makecell{\textbf{Cross-Entropy}} \\ \hline
developer\_accepted\_ratio    & 0.065          & -0.898          & 0.039          & -0.541          \\ \hline
developer\_accepted\_counts   & 0.006          & -0.082          & 0.004          & -0.049          \\ \hline
developer\_generated\_interval& 0.003          & -0.045          & 0.004          & -0.055          \\ \hline
developer\_generated\_lines   & 0.024          & -0.116          & 0.003          & -0.038          \\ \hline
developer\_generated\_chars   & 0.002          & -0.028          & 0.009          & -0.126          \\ \hline
developer\_presented\_time    & -0.001         & 0.017           & 0.000          & -0.005          \\ \hline
developer\_accepted\_time     & 0.014          & -0.187          & 0.002          & -0.022          \\ \hline
project\_accepted\_ratio      & 0.012          & -0.161          & 0.008          & -0.109          \\ \hline
project\_accepted\_counts     & 0.002          & -0.027          & 0.001          & -0.011          \\ \hline
project\_preceding\_lines     & 0.001          & -0.019          & 0.001          & -0.016          \\ \hline
project\_presented\_time      & -0.002         & 0.022           & 0.013          & -0.180          \\ \hline
in-situ\_IDE\_version         & \textbf{0.074} & \textbf{-1.029} & \textbf{0.096} & \textbf{-1.317} \\ \hline
\end{tabular}
\end{adjustbox}
\label{tab:individual_features}
\end{table}

To explore the weight of each feature in the \emph{CSAP} method and provide suggestions for model optimization, we conducted experiments removing each individual feature from the original CSAP method. Table~\ref{tab:individual_features} shows the difference between results with all features and results after removing each feature on the imbalanced test set. Due to space constraints, the table only shows the results for Accuracy and Cross-Entropy.
For Accuracy, positive values indicate that all features results outperform single-feature-removed results, and negative values indicate the opposite.
For Cross-Entropy, negative values indicate that single-feature-removed results perform worse.

Our individual feature analysis, detailed in Table~\ref{tab:individual_features}, reveals two key insights.

First, \textit{in-situ\_IDE\_version} is unequivocally the most critical feature. As shown by the bolded values, its removal triggers the most significant performance degradation across nearly all metrics on both test sets. On the imbalanced set, its absence reduces Accuracy by 0.074, and yields a Cross-Entropy delta of -1.029. The impact is similarly dominant on the balanced set, where Accuracy drops by 0.096 and the Cross-Entropy delta reaches -1.317. This finding shows that developers are sensitive to the modernity of code suggestions, making it crucial for models to stay synchronized with evolving code corpora.

Second, historical acceptance ratios are powerful predictors of future behavior. The \textit{developer\_accepted\_ratio} is particularly influential. Its removal leads to the second-largest drop in performance, decreasing accuracy by 0.065 and cross-entropy by -0.898 on the imbalanced set. This significant impact is mirrored on the balanced set, which sees a performance drop of 0.039 in accuracy and -0.541 in cross-entropy. Besides, \textit{project\_accepted\_ratio} also contributes meaningfully, though to a lesser extent. Together, these features highlight that the history of accepting or rejecting suggestions is a strong indicator of future behavior.

\bigskip
\begin{tcolorbox}
    \noindent\textbf{\emph{Answer to RQ\textsubscript{2}.}} We find that a personalized method to code acceptance prediction is highly effective. Our method, \emph{CSAP} significantly outperforms both an in-production industrial filter and a large language model baseline. Feature analysis reveals that the most critical predictors, stemming from in-situ context, and developer habits, are the \textit{in-situ\_IDE\_version}, and \textit{developer\_accepted\_ratio}. These features highlight the importance of tailoring predictions to a developer's immediate environment and past behavior.

\end{tcolorbox}
\medskip

\section{DISCUSSION}

\subsection{Implications}

Our findings offer several practical and research implications for the future of AI-assisted programming tools.

\textbf{The Case for Lightweight, Post-Generation Filtering.} 
Our findings indicate that a lightweight, post-generation filter provides the opportunity to improve the developer-perceived quality of an AI-assisted programming tool without modifying the underlying large language model. Specifically, our model \emph{CSAP} can be trained and updated frequently on fresh interaction data, making the system highly responsive to evolving developer and project dynamics in a way that is not feasible for large, computationally expensive foundation models. Our results suggest that a hybrid architecture—pairing a powerful general-purpose generator with a nimble, personalized filter—presents a practical and cost-effective path toward more adaptive and efficient AI-assisted development.

\textbf{Personalization as a First-Class Citizen.} 
Our findings establish that personalization is critical for code suggestion acceptance. The significance analysis in RQ1 revealed that historical context—particularly a developer's and project's past acceptance behaviors—significantly distinguishes accepted suggestions from rejected ones (Section~\ref{sec:rq1}). Building on this, our predictive model, \emph{CSAP}, confirmed these signals as powerful predictors (Section~\ref{sec:rq2}). Our feature contribution analysis reveals that historical acceptance ratios at both the developer and project levels are important. This evidence makes a strong case for treating personalization as a first-class citizen in AI-assisted programming. By learning from developer and project history, AI-assisted programming tools can evolve from generic generators to context-aware collaborators that provide more desired and less disruptive assistance.

\textbf{The Impact of Model Drift and Tool chain Evolution.} Our findings indicate that the evolution of the development tool chain has a significant influence on code acceptance. Specifically, we observed that the IDE version, which serves as a proxy for the underlying model and its training data vintage, is a highly significant predictor of acceptance (Section~\ref{sec:rq1}). This suggests that as development environments and their associated libraries evolve, the scope of code generation expands. If the underlying large language model is not specifically adjusted to account for these changes, its suggestions may become outdated or incompatible with new APIs and syntax—even when trained on up-to-date corpora—leading developers to reject them. Our results recommend that practitioners must keep models synchronized with evolving code corpora to prevent performance degradation from model drift.

\textbf{Quality Over Speed: The Interplay of Length and Latency.} 
Our findings challenge the conventional focus on minimizing latency, suggesting instead that developers prioritize suggestion quality. This is supported by a twofold analysis. First, the significance tests in RQ1 show that accepted suggestions are significantly longer and take more time to generate than rejected ones (Table~\ref{tab:overview of developer habit feature significance}, Table~\ref{tab:project_prefs_overview}). Second, our predictive model, \emph{CSAP}, confirms that historical suggestion length (\textit{developer\_generated\_lines}) and generation time for accepted suggestions (\textit{developer\_accepted\_time}) are informative signals, as their removal degrades performance (Table~\ref{tab:individual_features}). These results indicate that developers are willing to wait for more substantial, higher-quality generations. We therefore recommend that tool builders move beyond raw speed and instead implement adaptive latency budgets and learn personalized preferences for suggestion length to better align with developer intent.

\textbf{The Value of Fine-Grained Interaction Telemetry.} 
Our findings indicate that access to fine-grained interaction telemetry is a prerequisite for building the next generation of personalized AI programming assistants. Specifically, we observed that the significant features distinguishing accepted from rejected suggestions are historical features derived from detailed interaction logs (Section~\ref{sec:rq1}). These logs enable us to quantify nuanced developer- and project-specific patterns, such as historical acceptance ratios. Without telemetry capturing every accept and reject decision along with its surrounding context, these critical predictive signals would remain invisible. Our results suggest that organizations aiming to optimize AI-assisted development should treat interaction data as a first-class asset. This data-driven approach allows for a shift from one-size-fits-all code generation to a more collaborative and less disruptive paradigm, where AI assistance is tailored to the specific needs of the developer and the context of their work. 

\subsection{Threats to validity}

\textbf{External Validity.} 
Our study is conducted using telemetry data from a large technology company. The developer population, coding practices, project types, and software stack may not be representative of other industrial settings or the open-source community. 
Our dataset includes 66,239 developer-AI interaction logs from 12 projects,
and this large dataset reduces individual bias.
Due to the difficulty of collecting interaction data within company-internal environments,
this paper provides a first step to quantitatively analyze features and predict the acceptance of code suggestions,
which may inspire future work to explore more diverse development environments and confirm the broader applicability of our conclusions.

\textbf{Construct Validity.} 
Our features are proxies for complex underlying phenomena. For example, we use historical code metrics to represent a developer's "coding habits," but this is an incomplete view of their preferences and style. 
For example,
developers' experience in the company may also influence the acceptance of code suggestions.
In future work,
we hope to collect more information and explore their impacts on accepting or rejecting code suggestions.

\section{CONCLUSION}
\label{sec:conclusion}

AI-assisted programming tools are increasingly integrated into developer workflows, but their potential is often hindered by undesired suggestions that cause costly interruptions. In this study, we first conducted an empirical study on AI-developer interaction data, revealing that personal coding habits, the specific project preference, and the immediate properties of the code and development environment are significant features which distinguish accepted code suggestions and rejected ones. This finding underscores the necessity of personalization. Guided by these insights, we proposed \emph{CSAP}, a lightweight, personalized filter that predicts the acceptance of code suggestion before a suggestion is displayed. The evaluation results demonstrate that \emph{CSAP} achieves superior performance, significantly outperforming competitive baselines. To the best of our knowledge, it is the first quantitative study of code suggestion acceptance on large-scale industrial data, and this work also sheds light on an important research direction of AI-assisted programming.

\bibliographystyle{ACM-Reference-Format}
\bibliography{Reference}

@misc{tan2024far,
  author = {Tan, X. and Long, X. and Ni, X. and others},
  title = {How far are AI-powered programming assistants from meeting developers' needs?},
  year = {2024},
  eprint = {2404.12000},
  archivePrefix = {arXiv},
  primaryClass={cs.SE},
  url = {https://arxiv.org/abs/2404.12000}
}

@article{Fan2018Early,
  author = {Yuanrui Fan and· Xin Xia and· David Lo and·Shanping Li},
  title = {Early prediction of merged code changes to prioritize reviewing tasks},
  journal = {Empirical Software Engineering},
  volume = {23},
  pages = {3346–3393},
  year = {2018}
}

@article{barke2023grounded,
  author = {Barke, Shraddha and James, Michael B. and Polikarpova, Nadia},
  title = {Grounded Copilot: How Programmers Interact with Code-Generating Models},
  journal = {Proceedings of the ACM on Programming Languages},
  volume = {7},
  number = {OOPSLA1},
  pages = {85--111},
  year = {2023}
}

@inproceedings{liang2024large,
  author    = {Liang, J. T. and Yang, C. and Myers, B. A.},
  title     = {A large-scale survey on the usability of ai programming assistants: Successes and challenges},
  booktitle = {Proceedings of the 46th IEEE/ACM International Conference on Software Engineering(ICSE)},
  year      = {2024},
  pages     = {1--13},
  address   = {Lisbon, Portugal},
  publisher = {Association for Computing Machinery}
}

@misc{sun2024dontcompleteitpreventing,
      title={Don't Complete It! Preventing Unhelpful Code Completion for Productive and Sustainable Neural Code Completion Systems}, 
      author={Zhensu Sun and Xiaoning Du and Fu Song and Shangwen Wang and Mingze Ni and Li Li and David Lo},
      year={2024},
      eprint={2209.05948},
      archivePrefix={arXiv},
      primaryClass={cs.SE},
      url={https://arxiv.org/abs/2209.05948}, 
}

@inproceedings{azeem2020action,
  title={Action-based recommendation in pull-request development},
  author={Azeem, Muhammad Ilyas and Panichella, Sebastiano and Di Sorbo, Andrea and Serebrenik, Alexander and Wang, Qing},
  booktitle={Proceedings of the International Conference on Software and System Processes},
  pages={115--124},
  year={2020}
}

@inproceedings{joshi2024comparative,
  title={Comparative Study of Reinforcement Learning in GitHub Pull Request Outcome Predictions},
  author={Joshi, Rinkesh and Kahani, Nafiseh},
  booktitle={2024 IEEE International Conference on Software Analysis, Evolution and Reengineering (SANER)},
  pages={489--500},
  year={2024},
  organization={IEEE}
}

@article{islam2022early,
  title={Early prediction for merged vs abandoned code changes in modern code reviews},
  author={Islam, Khairul and Ahmed, Toufique and Shahriyar, Rifat and Iqbal, Anindya and Uddin, Gias},
  journal={Information and Software Technology},
  volume={142},
  pages={106756},
  year={2022},
  publisher={Elsevier}
}

@article{khatoonabadi2023wasted,
  title={On wasted contributions: Understanding the dynamics of contributor-abandoned pull requests--a mixed-methods study of 10 large open-source projects},
  author={Khatoonabadi, SayedHassan and Costa, Diego Elias and Abdalkareem, Rabe and Shihab, Emad},
  journal={ACM Transactions on Software Engineering and Methodology},
  volume={32},
  number={1},
  pages={1--39},
  year={2023},
  publisher={ACM New York, NY}
}

@article{zhang2022pull,
  title={Pull request decisions explained: An empirical overview},
  author={Zhang, Xunhui and Yu, Yue and Gousios, Georgios and Rastogi, Ayushi},
  journal={IEEE Transactions on Software Engineering},
  volume={49},
  number={2},
  pages={849--871},
  year={2022},
  publisher={IEEE}
}

@inproceedings{wang2021accept,
  title={Accept or not? an empirical study on analyzing the factors that affect the outcomes of modern code review?},
  author={Wang, Dandan and Wang, Qing and Wang, Junjie and Shi, Lin},
  booktitle={2021 IEEE 21st International Conference on Software Quality, Reliability and Security (QRS)},
  pages={946--955},
  year={2021},
  organization={IEEE}
}

@inproceedings{ciniselli2023source,
  title={Source code recommender systems: The practitioners' perspective},
  author={Ciniselli, Matteo and Pascarella, Luca and Aghajani, Emad and Scalabrino, Simone and Oliveto, Rocco and Bavota, Gabriele},
  booktitle={2023 IEEE/ACM 45th International Conference on Software Engineering (ICSE)},
  pages={2161--2172},
  year={2023},
  organization={IEEE}
}

@inproceedings{ross2023programmer,
  title={The programmer’s assistant: Conversational interaction with a large language model for software development},
  author={Ross, Steven I and Martinez, Fernando and Houde, Stephanie and Muller, Michael and Weisz, Justin D},
  booktitle={Proceedings of the 28th International Conference on Intelligent User Interfaces},
  pages={491--514},
  year={2023}
}

@inproceedings{ziegler2022productivity,
  title={Productivity assessment of neural code completion},
  author={Ziegler, Albert and Kalliamvakou, Eirini and Li, X Alice and Rice, Andrew and Rifkin, Devon and Simister, Shawn and Sittampalam, Ganesh and Aftandilian, Edward},
  booktitle={Proceedings of the 6th ACM SIGPLAN International Symposium on Machine Programming},
  pages={21--29},
  year={2022}
}

@inproceedings{zheng2023codegeex,
  title={Codegeex: A pre-trained model for code generation with multilingual benchmarking on humaneval-x},
  author={Zheng, Qinkai and Xia, Xiao and Zou, Xu and Dong, Yuxiao and Wang, Shan and Xue, Yufei and Shen, Lei and Wang, Zihan and Wang, Andi and Li, Yang and others},
  booktitle={Proceedings of the 29th ACM SIGKDD Conference on Knowledge Discovery and Data Mining},
  pages={5673--5684},
  year={2023}
}

@article{mann1947test,
  title={On a test of whether one of two random variables is stochastically larger than the other},
  author={Mann, Henry B and Whitney, Donald R},
  journal={The annals of mathematical statistics},
  pages={50--60},
  year={1947},
  publisher={JSTOR}
}

@inproceedings{vaithilingam2022expectation,
  title={Expectation vs. experience: Evaluating the usability of code generation tools powered by large language models},
  author={Vaithilingam, Priyan and Zhang, Tianyi and Glassman, Elena L},
  booktitle={Chi conference on human factors in computing systems extended abstracts},
  pages={1--7},
  year={2022}
}

@article{pearce2025asleep,
  title={Asleep at the keyboard? assessing the security of github copilot’s code contributions},
  author={Pearce, Hammond and Ahmad, Baleegh and Tan, Benjamin and Dolan-Gavitt, Brendan and Karri, Ramesh},
  journal={Communications of the ACM},
  volume={68},
  number={2},
  pages={96--105},
  year={2025},
  publisher={ACM New York, NY, USA}
}

@inproceedings{tu2014localness,
  title={On the localness of software},
  author={Tu, Zhaopeng and Su, Zhendong and Devanbu, Premkumar},
  booktitle={Proceedings of the 22nd ACM SIGSOFT international symposium on foundations of software engineering},
  pages={269--280},
  year={2014}
}

@article{levenshtein1966binary,
  title={Binary codes capable of correcting deletions, insertions, and reversals},
  author={Levenshtein, Vladimir Iosifovich},
  journal={Soviet Physics Doklady},
  volume={10},
  number={8},
  pages={707--710},
  year={1966}
}

@inproceedings{davila2024industry,
  title={An industry case study on adoption of ai-based programming assistants},
  author={Davila, Nicole and Wiese, Igor and Steinmacher, Igor and Lucio da Silva, Lucas and Kawamoto, Andr{\'e} and Favaro, Gilson Jos{\'e} Peres and Nunes, Ingrid},
  booktitle={Proceedings of the 46th International Conference on Software Engineering: Software Engineering in Practice},
  pages={92--102},
  year={2024}
}

@inproceedings{erhabor2025measuring,
  title={Measuring the Runtime Performance of C++ Code Written by Humans using GitHub Copilot},
  author={Erhabor, Daniel and Udayashankar, Sreeharsha and Nagappan, Meiyappan and Al-Kiswany, Samer},
  booktitle={2025 IEEE/ACM 47th International Conference on Software Engineering (ICSE)},
  pages={596--596},
  year={2025},
  organization={IEEE Computer Society}
}

@article{wu2023personalized,
  title={Personalized news recommendation: Methods and challenges},
  author={Wu, Chuhan and Wu, Fangzhao and Huang, Yongfeng and Xie, Xing},
  journal={ACM Transactions on Information Systems},
  volume={41},
  number={1},
  pages={1--50},
  year={2023},
  publisher={ACM New York, NY}
}

@inproceedings{mozannar2024reading,
  title={Reading between the lines: Modeling user behavior and costs in AI-assisted programming},
  author={Mozannar, Hussein and Bansal, Gagan and Fourney, Adam and Horvitz, Eric},
  booktitle={Proceedings of the 2024 CHI Conference on Human Factors in Computing Systems},
  pages={1--16},
  year={2024}
}

@inproceedings{bao2017will,
  title={Who will leave the company?: a large-scale industry study of developer turnover by mining monthly work report},
  author={Bao, Lingfeng and Xing, Zhenchang and Xia, Xin and Lo, David and Li, Shanping},
  booktitle={2017 IEEE/ACM 14th International Conference on Mining Software Repositories (MSR)},
  pages={170--181},
  year={2017},
  organization={IEEE}
}

@article{kamei2012large,
  title={A large-scale empirical study of just-in-time quality assurance},
  author={Kamei, Yasutaka and Shihab, Emad and Adams, Bram and Hassan, Ahmed E and Mockus, Audris and Sinha, Anand and Ubayashi, Naoyasu},
  journal={IEEE Transactions on Software Engineering},
  volume={39},
  number={6},
  pages={757--773},
  year={2012},
  publisher={IEEE}
}

@inproceedings{nagappan2005use,
  title={Use of relative code churn measures to predict system defect density},
  author={Nagappan, Nachiappan and Ball, Thomas},
  booktitle={Proceedings of the 27th international conference on Software engineering},
  pages={284--292},
  year={2005}
}

@article{guo2025deepseek,
  title={Deepseek-r1: Incentivizing reasoning capability in llms via reinforcement learning},
  author={Guo, Daya and Yang, Dejian and Zhang, Haowei and Song, Junxiao and Zhang, Ruoyu and Xu, Runxin and Zhu, Qihao and Ma, Shirong and Wang, Peiyi and Bi, Xiao and others},
  journal={arXiv preprint arXiv:2501.12948},
  year={2025}
}

@article{yang2025qwen3,
  title={Qwen3 technical report},
  author={Yang, An and Li, Anfeng and Yang, Baosong and Zhang, Beichen and Hui, Binyuan and Zheng, Bo and Yu, Bowen and Gao, Chang and Huang, Chengen and Lv, Chenxu and others},
  journal={arXiv preprint arXiv:2505.09388},
  year={2025}
}

@article{comanici2025gemini,
  title={Gemini 2.5: Pushing the frontier with advanced reasoning, multimodality, long context, and next generation agentic capabilities},
  author={Comanici, Gheorghe and Bieber, Eric and Schaekermann, Mike and Pasupat, Ice and Sachdeva, Noveen and Dhillon, Inderjit and Blistein, Marcel and Ram, Ori and Zhang, Dan and Rosen, Evan and others},
  journal={arXiv preprint arXiv:2507.06261},
  year={2025}
}

@article{li2022competition,
  title={Competition-level code generation with alphacode},
  author={Li, Yujia and Choi, David and Chung, Junyoung and Kushman, Nate and Schrittwieser, Julian and Leblond, R{\'e}mi and Eccles, Tom and Keeling, James and Gimeno, Felix and Dal Lago, Agustin and others},
  journal={Science},
  volume={378},
  number={6624},
  pages={1092--1097},
  year={2022},
  publisher={American Association for the Advancement of Science}
}

@article{gao2025trae,
  title={Trae Agent: An LLM-based Agent for Software Engineering with Test-time Scaling},
  author={Gao, Pengfei and Tian, Zhao and Meng, Xiangxin and Wang, Xinchen and Hu, Ruida and Xiao, Yuanan and Liu, Yizhou and Zhang, Zhao and Chen, Junjie and Gao, Cuiyun and others},
  journal={arXiv preprint arXiv:2507.23370},
  year={2025}
}

@inproceedings{wang2023practitioners,
  title={How practitioners expect code completion?},
  author={Wang, Chaozheng and Hu, Junhao and Gao, Cuiyun and Jin, Yu and Xie, Tao and Huang, Hailiang and Lei, Zhenyu and Deng, Yuetang},
  booktitle={Proceedings of the 31st ACM Joint European Software Engineering Conference and Symposium on the Foundations of Software Engineering},
  pages={1294--1306},
  year={2023}
}

@inproceedings{kyaw2018proposal,
  title={A proposal of code completion problem for java programming learning assistant system},
  author={Kyaw, Htoo Htoo Sandi and Aung, Shwe Thinzar and Thant, Hnin Aye and Funabiki, Nobuo},
  booktitle={Conference on Complex, Intelligent, and Software Intensive Systems},
  pages={855--864},
  year={2018},
  organization={Springer}
}

@article{bird2022taking,
  title={Taking Flight with Copilot: Early insights and opportunities of AI-powered pair-programming tools},
  author={Bird, Christian and Ford, Denae and Zimmermann, Thomas and Forsgren, Nicole and Kalliamvakou, Eirini and Lowdermilk, Travis and Gazit, Idan},
  journal={Queue},
  volume={20},
  number={6},
  pages={35--57},
  year={2022},
  publisher={ACM New York, NY, USA}
}

@misc{anthropic2024claude3_7,
  title   = {Claude 3.7 Sonnet and Claude Code},
  author  = {{Anthropic}},
  year    = {2024},
  month   = {October},
  howpublished = {\url{https://www.anthropic.com/news/claude-3-7-sonnet}},
  note    = {Accessed: 2025-09-05}
}

@article{bakal2025experience,
  title={Experience with GitHub Copilot for Developer Productivity at Zoominfo},
  author={Bakal, Gal and Dasdan, Ali and Katz, Yaniv and Kaufman, Michael and Levin, Guy},
  journal={arXiv preprint arXiv:2501.13282},
  year={2025}
}

@article{zhou2025exploring,
  title={Exploring the problems, their causes and solutions of AI pair programming: A study on GitHub and Stack Overflow},
  author={Zhou, Xiyu and Liang, Peng and Zhang, Beiqi and Li, Zengyang and Ahmad, Aakash and Shahin, Mojtaba and Waseem, Muhammad},
  journal={Journal of Systems and Software},
  volume={219},
  pages={112204},
  year={2025},
  publisher={Elsevier}
}

@misc{dresselhaus2025field,
  title = {Field Report: Coding in the Age of AI with Cursor},
  author = {Dresselhaus, Nicole},
  year = {2025},
  url = {https://drezil.de/Writing/coding-age-ai.html},
  urldate = {2025-09-07}
}

@article{guglielmi2025copilot,
  title={How do Copilot Suggestions Impact Developers' Frustration and Productivity?},
  author={Guglielmi, Emanuela and Arnoudova, Venera and Bavota, Gabriele and Oliveto, Rocco and Scalabrino, Simone},
  journal={arXiv preprint arXiv:2504.06808},
  year={2025}
}

@article{simkute2025ironies,
  title={Ironies of generative AI: understanding and mitigating productivity loss in Human-AI interaction},
  author={Simkute, Auste and Tankelevitch, Lev and Kewenig, Viktor and Scott, Ava Elizabeth and Sellen, Abigail and Rintel, Sean},
  journal={International Journal of Human--Computer Interaction},
  volume={41},
  number={5},
  pages={2898--2919},
  year={2025},
  publisher={Taylor \& Francis}
}

@inproceedings{hellendoorn2019code,
  title={When code completion fails: A case study on real-world completions},
  author={Hellendoorn, Vincent J and Proksch, Sebastian and Gall, Harald C and Bacchelli, Alberto},
  booktitle={2019 IEEE/ACM 41st International Conference on Software Engineering (ICSE)},
  pages={960--970},
  year={2019},
  organization={IEEE}
}

@article{pandey2024transforming,
  title={Transforming software development: Evaluating the efficiency and challenges of github copilot in real-world projects},
  author={Pandey, Ruchika and Singh, Prabhat and Wei, Raymond and Shankar, Shaila},
  journal={arXiv preprint arXiv:2406.17910},
  year={2024}
}

@inproceedings{
  loshchilov2019decoupled,
  title={Decoupled Weight Decay Regularization},
  author={Ilya Loshchilov and Frank Hutter},
  booktitle={International Conference on Learning Representations (ICLR)},
  year={2019},
  url={https://openreview.net/forum?id=Bkg6UKwA-}
}

@inproceedings{brown2024identifying,
  title={Identifying the factors that influence trust in AI code completion},
  author={Brown, Adam and D'Angelo, Sarah and Murillo, Ambar and Jaspan, Ciera and Green, Collin},
  booktitle={Proceedings of the 1st ACM International Conference on AI-Powered Software},
  pages={1--9},
  year={2024}
}

@inproceedings{li2021toward,
  title={Toward less hidden cost of code completion with acceptance and ranking models},
  author={Li, Jingxuan and Huang, Rui and Li, Wei and Yao, Kai and Tan, Weiguo},
  booktitle={2021 IEEE International Conference on Software Maintenance and Evolution (ICSME)},
  pages={195--205},
  year={2021},
  organization={IEEE}
}

@article{suryavanshi2025integrating,
  title={Integrating ChatGPT into Software Development: Valuating Acceptance and Utilisation Among Developers},
  author={Suryavanshi, Prathmesh and Kapse, Manohar and Sharma, Vinod},
  journal={Australasian Accounting, Business and Finance Journal},
  volume={19},
  number={1},
  year={2025}
}
\end{document}